\def\hunits{\hbox{\(\rm \,km\,s^{-1}\,Mpc^{-1}\)}}

\documentstyle[11pt,aaspp]{article}
\begin{document}

\title{The Effects of Starbursts and Low-Surface-Brightness Galaxies
on Faint Galaxy Models}
\author{B. A. McLeod}
\affil{Harvard-Smithsonian Center for Astrophysics,
60 Garden Street, MS-20, Cambridge, MA 02138\\
Email: bmcleod@cfa.harvard.edu}
\and
\author {M. J. Rieke}
\affil{Steward Observatory, University of Arizona, Tucson, AZ 85721\\
Email: mrieke@as.arizona.edu}

\begin{abstract}
We present models predicting the magnitude, redshift, and color
distributions of field galaxies.  We explore whether a fading
starburst scenario can account for the observations of faint blue
galaxies.  We marginally rule out a starburst scenario with a local
IMF because the models predict too many nearby faint galaxies that are
not observed. A burst model with a truncated IMF reproduces the counts
and redshift distributions well but produces too blue a population.
We show that surface brightness selection has a significant effect on
the distributions.  In particular, adding a population of low-surface
brightness galaxies, known to exist locally, can explain the counts
for $B_J<23$. They do not, however, account for the steepness of the
counts at fainter magnitudes.  Finally we show that the colors of
galaxies in a $K$-band selected sample are too blue to be consistent with a 
constant star-formation rate, passive-evolution model.
\end{abstract}

\section{Introduction}
A powerful test of our understanding of galaxy evolution and cosmology
lies in understanding the observed properties of galaxies over as
large a range of wavelengths and magnitudes as possible.  We wish to
create a model which specifies the basic cosmological parameters
($q_0$ and $H_0$) and a description of the distribution of
the physical properties of galaxies, i.e., their spectra,
luminosities, and sizes.  A complete description must also include the
time-dependence of these properties.  From this description we can
then predict observable properties: the distribution of apparent
magnitudes, colors, and redshifts.  Since we cannot directly observe
the evolution of individual galaxies, our understanding of evolution
can be tested only through changes in the populations of galaxies.

Hubble (1936) showed that the number of galaxies per magnitude
interval, $n(m)$, followed the relation $d\log n/dm = 0.6$ for bright
galaxies, as predicted by simple theory.  A detailed derivation
of $n(m)$ as a function of $q_0$ was done by Sandage (1961;1988), who showed
that, to first order, $n(m)$ is independent of $q_0$.
The importance of galaxy evolution was shown by
Brown \& Tinsley (1974)
who pointed out that ``a meaningful interpretation of the $N(m)$
relation will require good estimates of the luminosity functions,
spectral energy distributions and evolution of a wide range of types
of galaxies.''  However, the sophistication of the models to explain
the observed distributions remained relatively simple until recently,
partly due to a lack of observational constraints.

Early determinations (e.g. Peterson et al. 1979) showed that $n(m)$ in
the blue had a slope consistent with moderate amounts of luminosity
evolution.  The interpretation that a steep slope implies luminosity
evolution was made in the absence of redshift distributions, $n(z)$,
and it came as a surprise to many when the excess galaxies turned out
to be at fairly low redshifts (Broadhurst, Ellis, \& Shanks 1988).  An
additional puzzle came with deep $K$-band field counts
(\cite{Cowie91}; Gardner, Cowie, \& Wainscoat 1993), which showed no
excess of galaxies over that predicted by a simple no-evolution model.

There have been several approaches to explaining the apparent
discrepancy between the simple models and the observations.  One
possibility is the addition of a population of starbursting galaxies
(Lilly, Cowie, \& Gardner 1991) with the number of bursting galaxies
increasing with redshift.  These galaxies have since faded, been
disrupted or merged.  Broadhurst, Ellis, \& Glazebrook (1992)
addressed the merging scenario with a model that combined merging with
substantial luminosity evolution to match the $K$ and $B_J$ counts and
$B_J$ redshift distributions.  However, it is not clear that the
present population of galaxies can account for the required merger
products (Dalcanton 1993).  Yoshii (1993) has recently argued that
introducing a non-zero cosmological constant is sufficient to bring
the models into agreement.  On the other hand, gravitational lensing
statistics severely constrain the cosmological constant (Maoz \& Rix
1993).  Koo \& Kron (1992) and Koo, Gronwall, \& Bruzual (1993) have
argued that the local luminosity function is not sufficiently well
known to rule out models with no merging and no new populations.  They
present a no-evolution model that fits a large range of observables,
but at the expense of removing nearly all the luminous red galaxies
from the local luminosity function.  Continuing with this strategy,
Gronwall \& Koo (1995) have added passive evolution to the models and 
further improved the quality of their fits.  However, the starbursting 
galaxies in their models are still treated in an unphysical, no-evolution 
manner.

With such a complex problem, only a limited set of scenarios can be
considered.  Here we restrict ourselves to zero cosmological constant
and we do not consider galaxy merging. We concentrate on evaluating
the effects of adding a fading starburst population and considering
surface-brightness selection effects.  The evolutionary properties of
the galaxies, including starbursts, are constrained by
stellar-evolution models, i.e.  no arbitrary luminosity evolution of
galaxies will be allowed (e.g. Lilly 1993).  We also try to evaluate
each model with a large number of observational constraints.

We begin with a brief description of the distributions to be computed.
In \S~\ref{sec-data} we present a compilation of previous
observational results.  Section~\ref{sec-quiesc} considers a simple
model with a quiescently evolving population.  The next section
introduces starburst populations to examine their effects on the
excess blue counts. This is followed by a discussion of surface
brightness effects in \S~\ref{sec-surface} including the effects of
known low-surface-brightness galaxies. We conclude with the implications
that the $I-K$ colors of a $K$-selected sample have on evolution around $z$=1.

\section{Theory}
We consider three basic types of distributions: 1) number of galaxies as a
function of apparent magnitude for a given filter; 2) number of
galaxies as a function of redshift in a particular apparent magnitude
range; and 3) number of galaxies as a function of color in a
particular apparent magnitude range. The first quantity, the
number-magnitude relation, is given by 
\begin{equation} n(m) = \sum_i
\int_0^{z_{f}} \phi_i(M(m,z)) {dV\over{dz}} dz.
\label{nm-eq}\end{equation}
 The sum is over galaxies of different spectral types, $\phi_i$ is the
luminosity function for galaxy type $i$, $z_f$ is the redshift the
galaxies formed, and the volume element, $dV/dz$, is determined by the
cosmology.  The relation between apparent and absolute magnitude is
given by
\begin{equation} M = m - 5 \log
{d_L(z)\over 10\,\rm{pc}} - 2.5 \log {(1+z)\int_0^\infty
S_i(\lambda,t(z=0))F(\lambda)d\lambda \over
\int_0^\infty S_i({\lambda\over 1+z},t(z))F(\lambda)d\lambda} - 
\Delta_i(M,z),\label{mag-eq} \end{equation} where $d_L$
is the luminosity distance, $S$ is the spectrum of the galaxy as a
function of time, $F$ is the filter transmission, and $\Delta$ is the
aperture correction. 
The third term, the KE correction, is due to two 
effects: the redshift of the galaxy shifts the observed bandpass to a bluer
and narrower range of
emitted wavelengths ($k$-correction), and the spectrum of the galaxy changes with time (evolution).
For now we will assume that $\Delta=0$, i.e.,
the observed magnitude measures the total flux from the galaxy. 
This
will be discussed further below.

The number of galaxies in a given redshift and magnitude bin is
\begin{equation}
n(z_l,z_u,m_l,m_u) = \sum_i \int_{z_l}^{z_u} \int_{m_l}^{m_u} 
\phi_i(M(m,z)) dm {dV\over{dz}} dz.\label{nz-eq} \end{equation}
The color distributions are determined by performing the integrals
over magnitude and redshift, but at each step of the integration the
color of the galaxy is determined and the luminosity function value is
added to the appropriate color bin.

\section{Data}
\label{sec-data}
Numerous observations of galaxy count vs. magnitude have been made.
To obtain the best estimate of the number of galaxies at each apparent
magnitude, we have combined the observations of several authors.  Each
data point is an average of the available data, weighted by the area
surveyed.  We present data only where the total number of galaxies
observed is greater than 20. Data for $U$, $B_J$ ($B_J=B-0.3(B-V)$),
Gunn-$r$, $I$, and $K$ are presented in Tables 1--5, respectively.
The tables list the centers of the magnitude bins, the number of
galaxies per square degree per magnitude, and the number of galaxies
actually observed in each bin.  The sources are drawn from the lists
in the recent review by Koo \& Kron (1992) and the $K$-band summary by
Gardner et al. (1993).  Additional $K$ data are from McLeod et
al. (1995).  In some cases the data to be combined were made through
somewhat different filters so we have applied the color
transformations given by the authors.  Note that the zero point for
$r$ is based on an F-star spectrum (Thuan \& Gunn 1976) and so differs
from the normal A0-star normalization in that $r$ is 0.43 mag larger
than expected.  The counts are from photographic surveys for $U<22$,
$B_J<23.5$, $r<22.0$, and $I<21.0$; fainter counts are from CCD
surveys.


\section{Quiescent Population}
\label{sec-quiesc}

\subsection{Spectral energy distributions}
\begin{figure}[tbp]
{\plotone{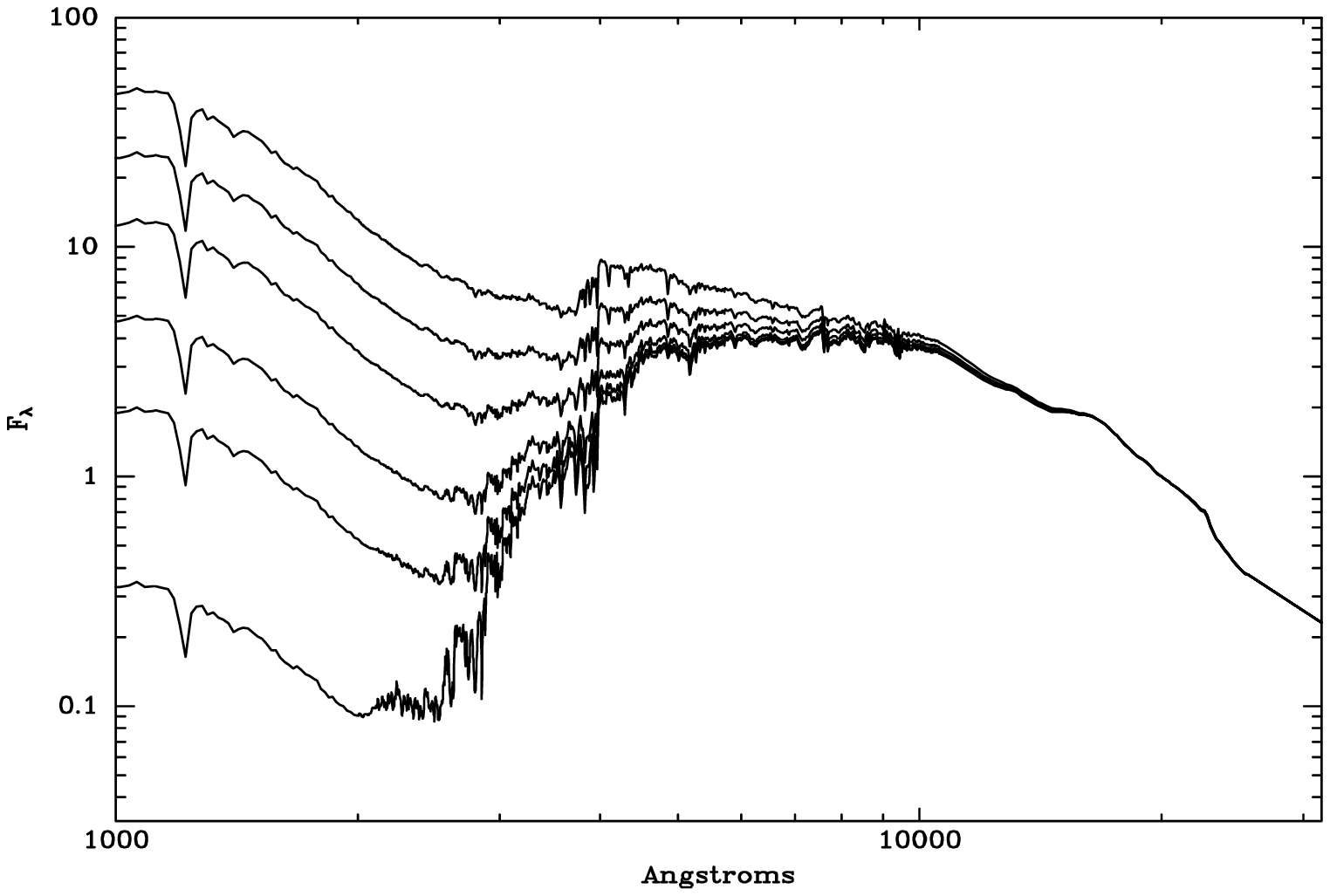}}
\caption[Model Spectral Energy Distributions]{\label{SEDs}}
Galaxy spectral energy distributions at an age of 13.5 Gyr generated
by Bruzual \& Charlot (1993) galaxy evolution models.  From top to
bottom, the constant star-formation rate per Gyr relative to the amount of
star formation in the initial burst
is 1.0, 0.03, 0.01, 0.003, 0.001, and 0.00001.
\end{figure}

\begin{figure}[tbp]
{\plotone{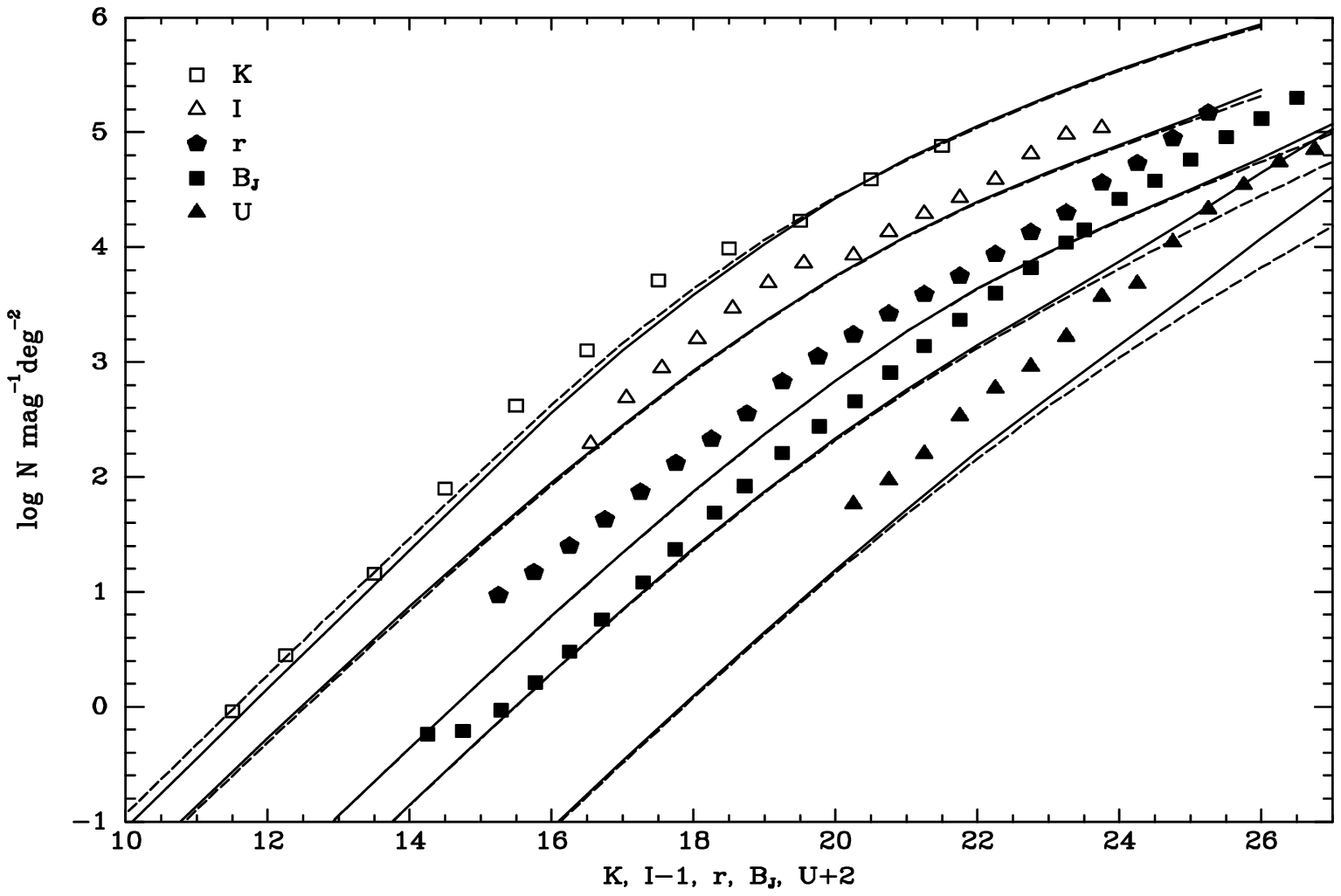}}
\caption[Comparison of non-evolving SEDs]{\label{Noevol}}
Comparison of expected counts using non-evolving
Bruzual SEDs (solid) vs Coleman et
al. SEDs (dashed). The biggest differences occur in the faint $B_J$ and $U$
counts where redshifted UV flux is detected.  Data plotted is listed in 
Tables~\ref{obsU-tab}--\ref{obsK-tab}.
\end{figure}

We use the evolution models of Bruzual \& Charlot (1993) to produce
galaxy spectral energy distributions (SEDs).  In all cases we have
assumed a Salpeter (1955) initial mass function (IMF),
i.e, $dn(m)/dm \propto m^{-2.35}$. Each of the models
consists of two components: an instantaneous burst at t=0 plus a
constant star formation rate (SFR) thereafter.  By adjusting the ratio of the
strength of the burst to the constant component we can affect the
shape of the SED.  The ratios were chosen so the models at an age of
13.5 Gyr would match the optical SEDs in Coleman, Wu, \& Weedman
(1980) with a near-infrared extension (Rieke \& Rieke, in
preparation) appended. Table~\ref{quiesc-lf}\ 
lists the amount of star formation per Gyr relative to the amount in the 
initial burst.
Figure~\ref{SEDs} shows these SEDs.
The early-type galaxies are completely dominated by the initial
burst; the late-type galaxies are dominated by the constant component.
Figure~\ref{Noevol}
shows a comparison of the expected counts using the Bruzual SEDs vs.
the Coleman et al. SEDs.  This Bruzual model assumes that there is no
evolution of the SEDs with time, an unphysical assumption, but
necessary for this purpose because the Coleman et al. SEDs contain no
evolutionary information. The purpose of this figure is to illustrate
the order of magnitude of the errors in the counts due to
uncertainties in the SEDs. Most of the divergence occurs in the faint
blue counts where the lack of understanding of the ultraviolet spectra of
galaxies becomes important. 

\subsection{Luminosity functions} 
\begin{figure}[tbp]
{\plotone{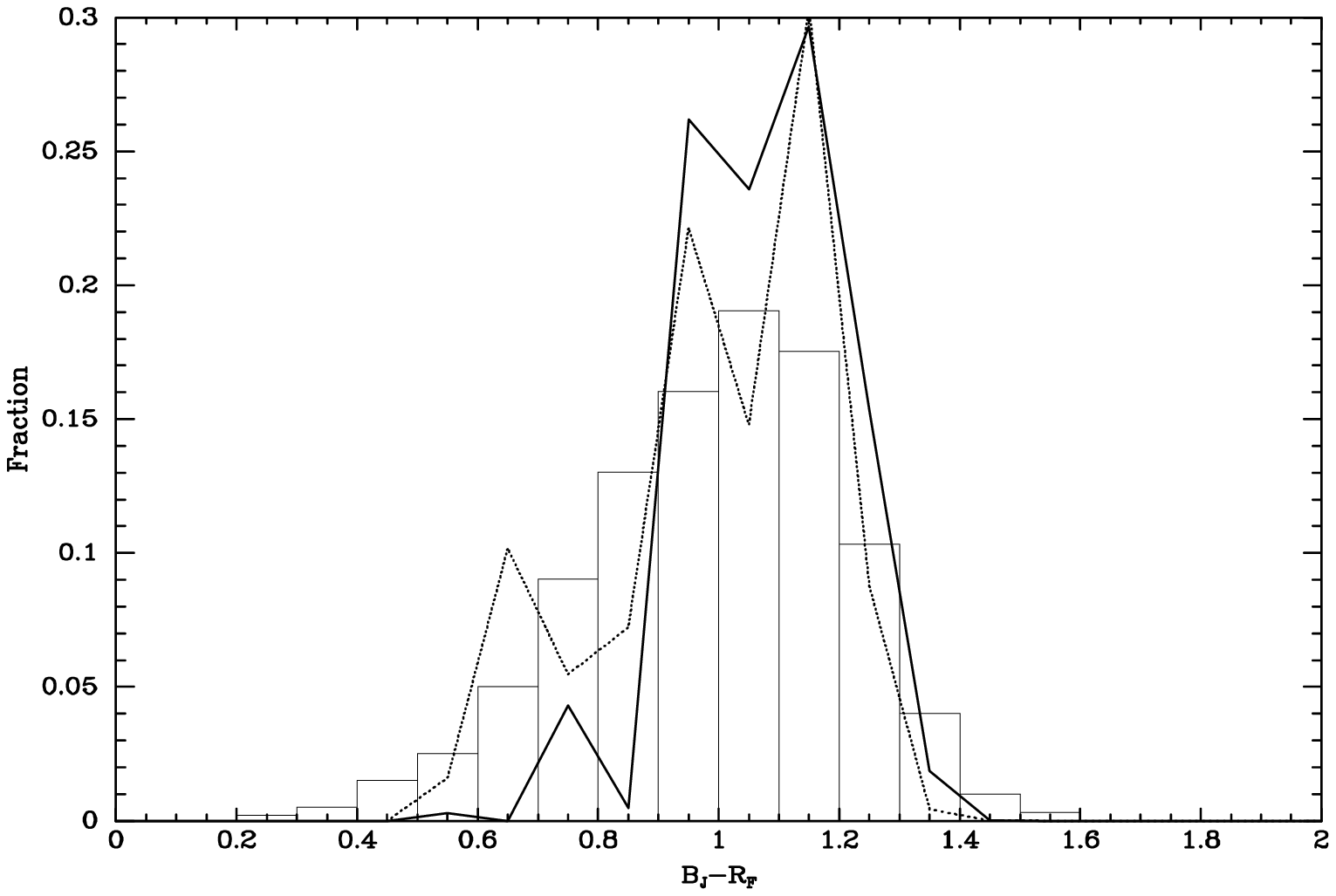}}
\caption[Color distribution for bright galaxies]{\label{BrightJF}}
Comparison of predicted color distributions ($B_J-R_F$) for bright galaxies
($15<B_J<17$) with observations. Transformations  are $B_J=B-0.3(B-V)$ and
$R_F=R-0.06(B-R)$.
 The histogram is observations from
Koo \& Kron (1992).  The solid line shows the Lilly (1993) type-dependent
LF. The dashed line shows the Shanks color-dependent LF.  We prefer
the Shanks LF because the colors match the observations better.
\end{figure}

We initially consider two different local luminosity functions (LFs).
The first set, used by Lilly (1993), are the type-dependent LFs
presented by Bingelli, Sandage, \& Tammann (1988), slightly adjusted
to sum to a Schechter function with parameters
$\phi^*=0.00175\rm\,Mpc^{-3}\,mag^{-1}$, $ M_{B_J}^* = -21.0$, and
$\alpha=-1.15$. The Schechter function (Schechter 1976) is defined by 
\( \phi(M) = 0.92 \phi^* \exp\{-0.92(M-M^*)(\alpha+1) -
\exp[-0.92(M-M^*)]\}\).  
This total LF is the one derived by Loveday et
al. (1992), but with a slightly steeper faint end slope.  All LF
specifications in this paper are scaled to $H_0=50\hunits$ for easy
comparison with other authors.  We pair each morphological type LF
with the corresponding Bruzual SED.  The second luminosity function we
consider is the color-dependent derivation by Shanks (1990).  This LF
is divided into three classes, $B-V<0.65$, $0.65<B-V<0.85$, and
$B-V>0.85$.  We further arbitrarily divide each class equally among
the SEDs which fall within that color range, as shown in
Table~\ref{quiesc-lf}.  The overall normalization of the LFs is chosen
so as to match the observed counts at $B_J$=17.  Figure~\ref{BrightJF}
shows a comparison of the resulting color
distributions of bright ($15<B_J<17$) galaxies  from the
two LFs along with the observed distribution presented in Koo \& Kron
(1992).  The Lilly LF produces too few blue galaxies relative to red
ones, while the Shanks LF has a more correct distribution. Ideally we
would like a LF with more than three divisions according to color, but
for now we will adopt the Shanks local LF.

\subsection{Results} 
\begin{figure}[tbp]
{\plotone{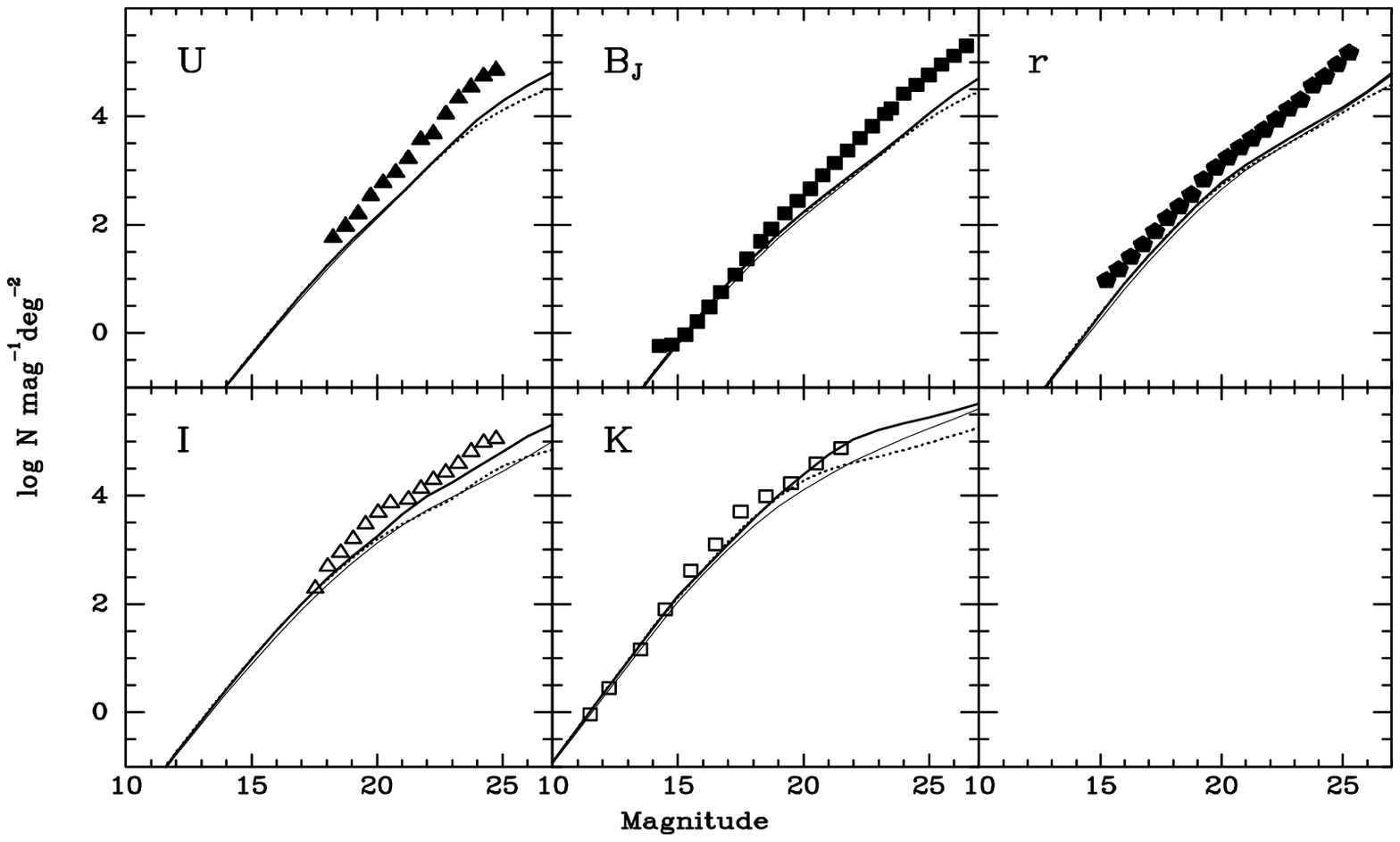}}
\caption[Galaxy counts for simple models]{\label{Base-nm}}
Galaxy numbers vs.\ magnitude for simple
models.  The thin line shows the unphysical non-evolving model.  The
solid line is the passively evolving population with $q_0$=0.05 and
$H_0=60\hunits$.  The dotted line is the same with $q_0$=0.5
and $H_0=47\hunits$.  The redshift of galaxy formation is z=7.3 in both
cases.
\end{figure}

In Figure~\ref{Base-nm} we show the
resulting counts through the $U$, $B_J$, $r$, $I$ and $K$ filters.
From above, we assume that the present day age of the quiescent
population is 13.5\,Gyr.  The value of $H_0$ was chosen so that the
galaxies are not formed at too low a redshift.  For example,
$H_0=50\hunits$ puts the redshift of formation at $z=2.5$, which would
be excluded by the redshift distributions (galaxies in their initial
burst would be seen at relatively bright magnitudes).  We have adopted a 
redshift of formation, 
$z_f$, of 7.3 and assume $q_0$=0.05, which leads to a choice of
$H_0=60\hunits$.  

 This baseline model (solid lines) shows the same basic feature of
galaxy count models made by various previous authors: the simple model
matches the $K$-band counts reasonably well, but drastically
underestimates the faint blue counts. All the models show a
significant discrepancy with the bright-end $r$ observations. The
reason for this is not clear. One possibility is that our SEDs are
somehow drastically wrong in the $r$ band. This seems unlikely since
the Coleman et al. SEDs in Figure~\ref{Noevol} produce the same
effect. Secondly, the observations could have a zero-point error; or
it could be a true large scale structure effect. Picard (1991) sees
$30\%$ differences between his northern and southern fields which he
attributes to structure.  His observations are a factor of two higher
than those of Sebok (1986).  Recent results from the second Palomar
Sky Survey (Weir, Djorgovski, \& Fayyad 1995) are intermediate
between Picard and Sebok.  These discrepancies will require further
work to resolve.

\begin{figure}[tbp]
{\plotone{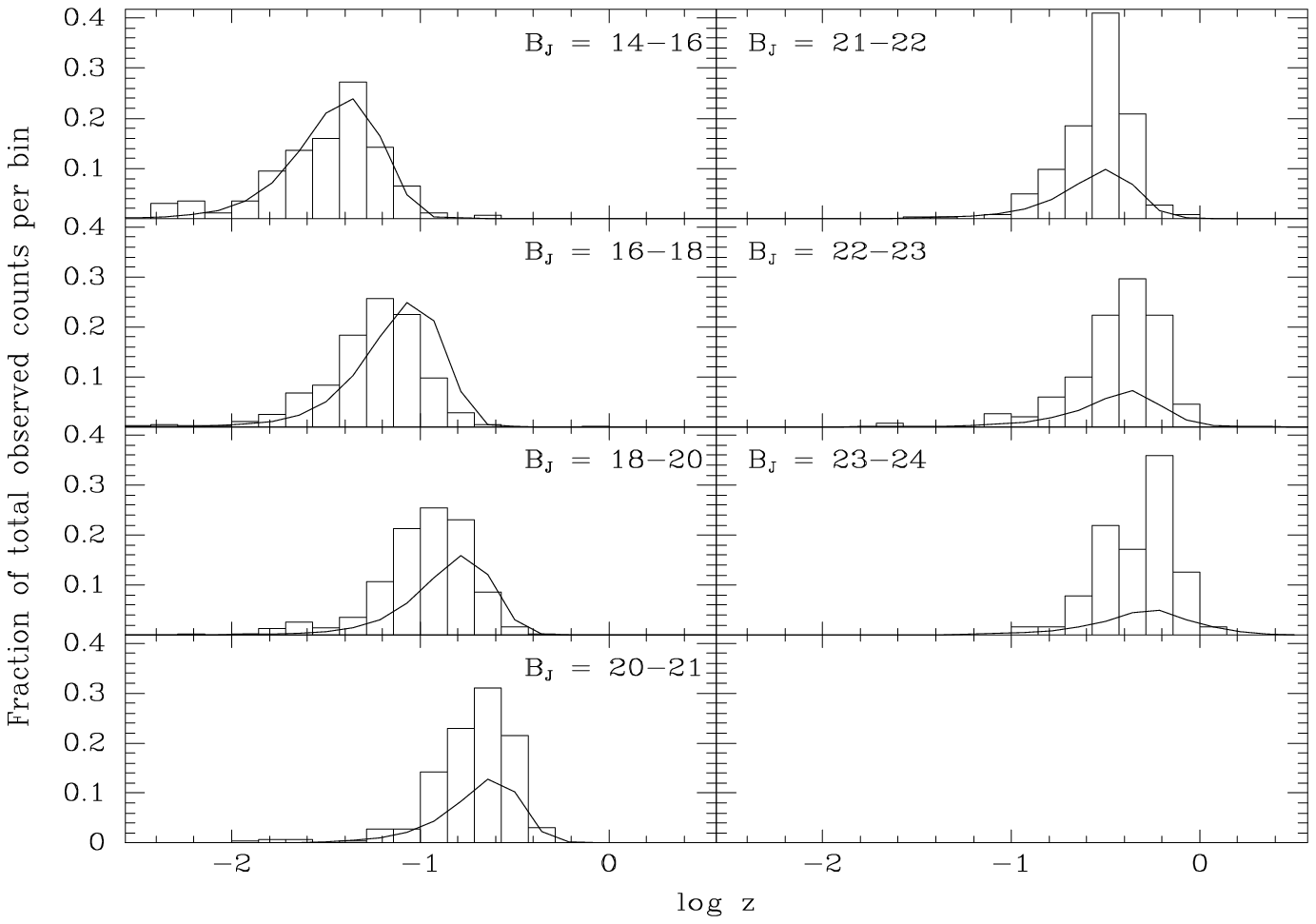}}
\caption[Redshift distributions for baseline model]{\label{Base-nz}}
Redshift distributions for the baseline passively evolving model
($q_0=0.05$, $H_0=60\hunits$, $t=13.5\,\rm Gyr$).
The solid line is the model prediction. The histograms are the
observations presented in Koo et al. (1993).  The models are not renormalized;
thus, the ratio of the number of predicted galaxies at each magnitude
relative to the number observed (Figure~\ref{Base-nm}) is preserved.
\end{figure}

 The dashed line of Figure~\ref{Base-nm} shows the same evolutionary model for
$q_0=0.5$. To preserve the present day SED ages of 13.5 Gyr with
$z_f$ = 7.3 we have changed $H_0$ to 47.  The only effect of
changing $H_0$ in these models is to change the relation of age to
redshift.  Figure~\ref{Base-nz} shows 
$B_J$-selected redshift distributions from Koo, Gronwall, \& Bruzual
(1993) with the predictions of the $q_0$=0.05 model superimposed.  As
in Koo et al. (1993), the histograms are shown in seven logarithmic
bins per decade.  Unlike other authors, we have normalized the
observed redshift distributions to match the observed total number of
galaxies in each magnitude interval rather than the predicted number.
The ratio of the area under the histogram to that under the model
curve in each panel should correspond to the prediction in the $n(m)$
plot.  Thus it is easy to see from the redshift distributions that the
model underpredicts the number of galaxies present for faint $B_J$.

\subsection{Alternate inputs}
Many different star formation histories can give rise to a given
present day SED.  We have also explored using exponential SFRs to
produce the SEDs and find that they produce too much luminosity
evolution. Specifically we used the exponential models adopted by
Gronwall \& Koo (1995), including reddening, combined with the
luminosity function adopted above. This exponential SFR model
overpredicts, by factors of several, the number of galaxies with
$B_J\approx22.5$ and $z>0.8$.  The model of Gronwall \& Koo has no
high redshift excess because their derived luminosity function
drastically reduces the number of bright galaxies of intermediate
spectral type compared with the derived LFs of Marzke, Huchra \& Geller
(1994) and of 
Shanks (1990).  We thus choose to use the burst-plus-constant form of
the SFR and limit the amount of luminosity evolution.  It is worth
remembering though that observationally derived LFs may have
significant errors at the extreme bright end, so this topic warrants
further investigation.  Exponential star-formation is revisited in
\S~\ref{sec-kcolors} when we discuss colors of $K-$selected galaxies.

We have also considered the effects of replacing our color-dependent
LF with the type-dependent LF of Marzke et al. (1994). This LF has a
more steeply rising faint end than the one we adopted.  We find that
as with the Lilly (1993) LF, we get a bright population that is
somewhat too red.  This is likely due to problems making a one-to-one
association of mophological types with color, and a more careful
treatment may give better results.  However, a comparison of $n(m)$
between the two LFs reveals that they give nearly identical results.
This gives us confidence that minor changes in the LF do not
dramatically change our results.

\section{Starburst Populations}
In this section we consider the effects of making up for the
deficiency of faint blue galaxies by adding a population of starburst
galaxies.  We assume that each new galaxy has a single burst of star
formation lasting 0.1 Gyr and thereafter fades quiescently.  This type
of population should provide the largest amount of fading possible and
test whether such galaxies can account for an excess of blue galaxies. 
For computational reasons we add the bursts at discrete points in time,
one burst population for each of the observed redshift distribution bins.

\begin{figure}[tbp]
{\plotone{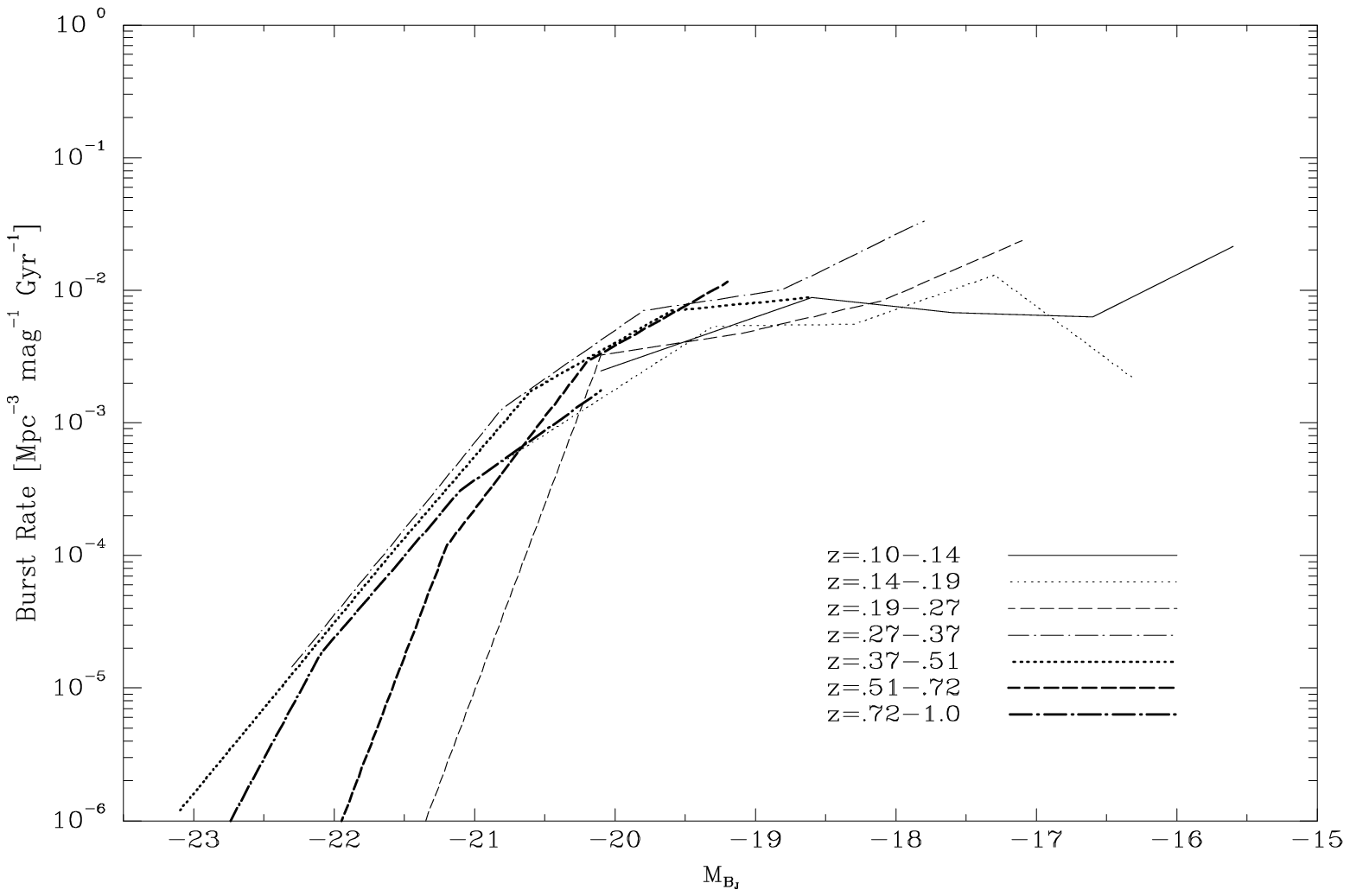}}
\caption[Derived burst luminosity functions]{\label{burstlf-fig}}
Derived burst rate luminosity functions for 0.1\,Gyr burst populations. The IMF
is Salpeter for 0.1-125$M_{\sun}$; $H_0=50\hunits$.  The horizontal axis
is the peak $B_J$ absolute magnitude.  The vertical axis is scaled in
terms of the number of galaxies that reach the specified peak
magnitude per unit comoving volume per Gyr.
\end{figure}

The luminosity functions of the burst galaxies are determined
empirically. By subtracting the baseline-model redshift distributions
from the observed distributions we get the number of excess galaxies
in each bin.  A further division by the comoving volume of the
redshift bin converts to a spatial density and one more normalization
corrects for the fact that the burst galaxies are in their bright
phase for only a portion of the time represented by the redshift bin
they are in. We assume an effective lifetime of the burst phase of 0.2
Gyr, i.e the time for the stars to form and fade by a factor of $e$.
The KE-correction and distance modulus determine the peak absolute
magnitudes required to produce the observed apparent magnitudes.  The
characteristics of the burst population are tabulated in
Table~\ref{burst-lf} and the derived burst rate LFs are shown in
Figure~\ref{burstlf-fig}.  There are no obvious trends of the derived
LFs as a function of redshift.  A fit done by eye shows that Schechter
parameters of $B_J^*=-19.0$,
$\phi^*=0.02\rm\,Mpc^{-3}\,mag^{-1}\,Gyr^{-1}$, and $\alpha=-1$ are a
reasonable fit to the ensemble of LFs. Here we have normalized the
amplitude $\phi^*$ in terms of the rate of burst formation per Gyr.
Note that since the absolute magnitudes are measured at the brightest
phase, a galaxy with $M_{B_J} = -19.0$ has a mass of
$2\times10^{8}\,M_{\sun}$, a factor of 100 less massive than an old
galaxy of the same luminosity would have.

\begin{figure}[tbp]
{\plotone{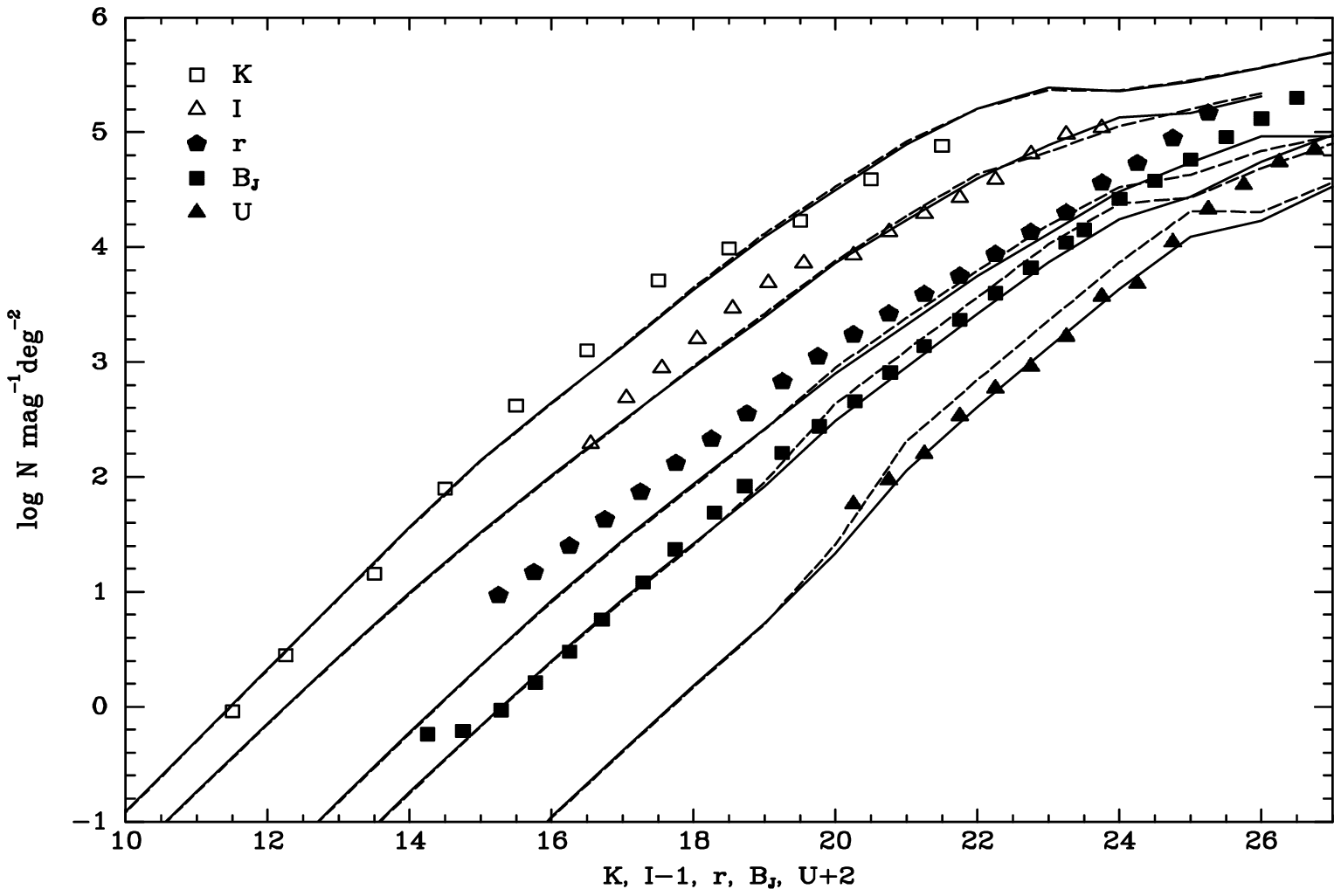}}
\caption[Number vs. magnitude for burst populations]{\label{Burst-nm}}
Number vs. magnitude for burst populations. Solid line is for
$0.1-125~M_{\sun}$ IMF.  Dashed line is for $2.5-125~M_{\sun}$ IMF.
\end{figure}

\begin{figure}[tbp]
{\plotone{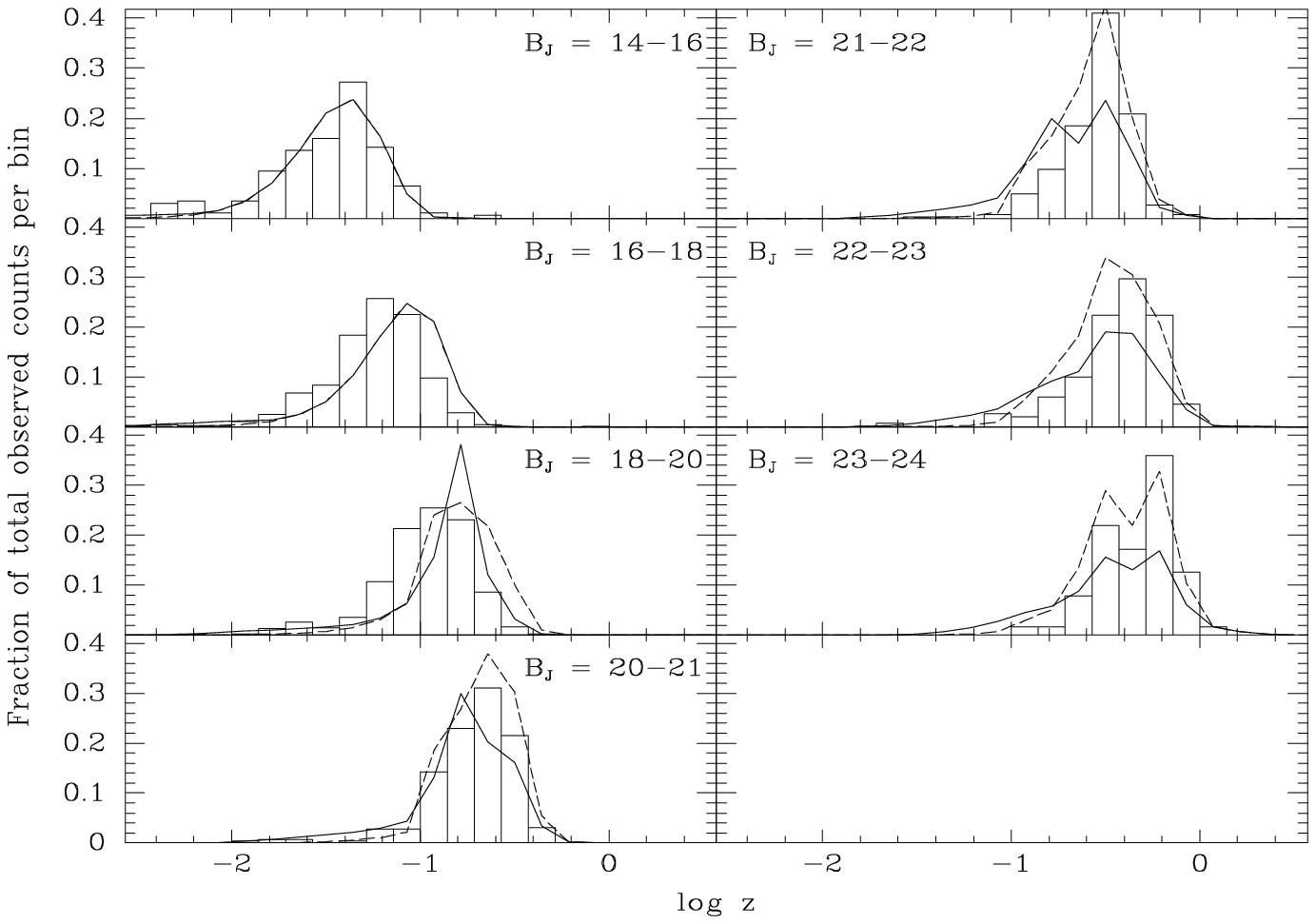}}
\caption[Redshift distributions for burst populations]{\label{Burst-nz}}
Redshift distributions for burst populations. Solid line is for
$0.1-125~M_{\sun}$ IMF.  Dashed line is for $2.5-125~M_{\sun}$ IMF.
\end{figure}

\begin{figure}[tbp]
{\plotone{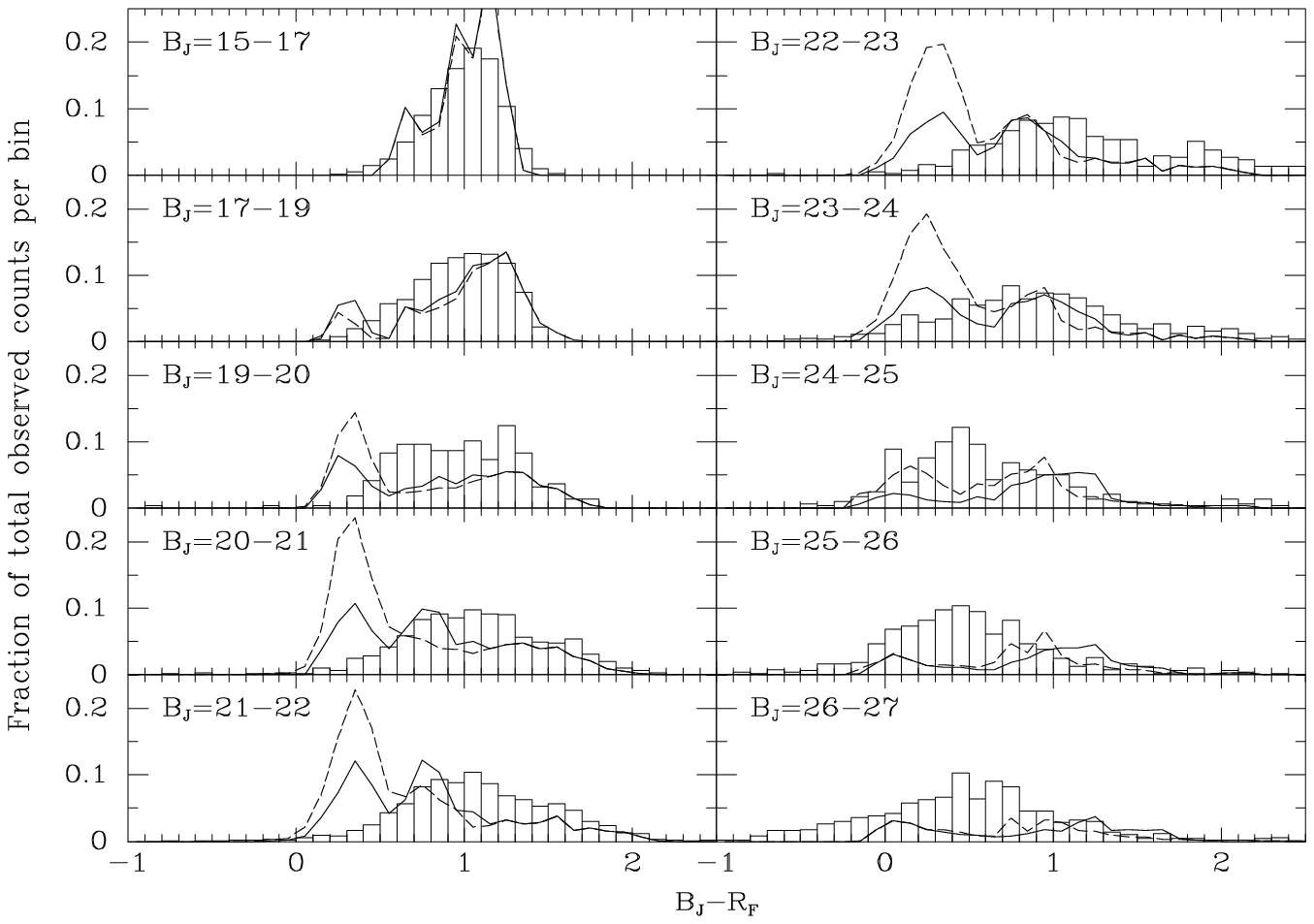}}
\caption[Color distributions for burst populations]{\label{Burst-nc}}
$B_J$--$R_F$ color distributions for burst populations. Solid line is for
$0.1-125~M_{\sun}$ IMF.  Dashed line is for $2.5-125~M_{\sun}$ IMF.
\end{figure}

Figures~\ref{Burst-nm}--\ref{Burst-nc} (solid lines) show the magnitude, redshift,
and color distributions for a burst model where the bursts have an IMF
identical to the baseline population (Salpeter, $0.1-125~M_{\sun}$).
Now the blue counts fit
reasonably well up to $B_J$=24. They turn over at fainter magnitudes because
we have no redshift distributions at fainter levels and have made no
attempt to extrapolate the burst population. 
The most serious concern is the redshift distributions.  Although the
high-$z$ end of each distribution fits well, this model predicts a low-$z$
tail in the $B_J>20$ distributions that is not seen in the
observations.  This effect is due to the fact that the low-mass stars
are long-lived and so the galaxy never fades completely.  This effect
is under-represented in this model since we have not added  $z>1$
bursts.  In \S~\ref{sec-surface} we will consider whether surface
brightness selection effects can reduce the low redshift tail.
One may also question how reliable the observations are.  Tresse et al. (1993)
find an excess of low-redshift galaxies in an $I-$band selected redshift 
survey.  However, this survey contains a small number of galaxies.
On the other hand, Glazebrook et al. (1995) argue that a low-redshift excess
is unlikely, based on their $22.5<B<24$ redshift survey.  They find no
excess over the baseline model.  These observational discrepancies will
become easier to resolve in the next few years with new multi-object 
spectrographs on large telescopes.

 The $B_J-R_F$ color distributions now show an excess at the blue end
but still a deficit at the red end.  This could be alleviated by
either adding reddening to the models or adding an older stellar
component to the galaxies.  However, both of these would exacerbate
the problem of the low-redshift tails.  Adding an older stellar
component would increase the amount of light that remains luminous to
the present.  Adding dust would require that, to maintain the same
observed blue luminosity at high redshift, we must increase the total
luminosity of the burst.  The net result would again be to decease the amount
of fading and increase the low-redshift population.  In reality, both dust
and old stars are likely to be present, and an improved model would take
these into account.

Modeling of the stellar populations of the nearby starburst galaxy M82
(Rieke et al. 1993) suggests that while the slope of the IMF in M82 may be
equal to that of the local IMF, very few low-mass stars are produced.
We now consider a starburst model using a Salpeter IMF but truncated
so that no stars with $M<2.5 M_{\sun}$ are produced.  An application
of the Bruzual \& Charlot (1993) models to this type of IMF is
presented by Charlot et al. (1993). The LFs derived are identical to
the ones above, but because the burst fades so rapidly, we assume that
the effective bright lifetime of the burst is the same as the burst
length, 0.1 Gyr. Thus the LFs have a normalization twice as large as
previously.  For this IMF a $M_{B_J}=-19.0$ galaxy has a mass of
$6\times10^7\,M_{\sun}$.  The amount of fading ranges from 15\,mag 
after 2\,Gyr to 19\,mag after 4.3\,Gyr.  After 5\,Gyr the galaxy
has faded completely because all the stars have turned into stellar
remnants.

Figures~\ref{Burst-nm}--\ref{Burst-nc} (dashed lines) show the
distributions for this model.  As in the previous burst model, the $U$
counts are over predicted.  The redshift distributions this time are
better, having no tails at low redshift.  The reason for this is
clear: the galaxies, having no low-mass stars, fade completely and so
are undetectable except in their burst phase.

\section{Surface Brightness Effects}
\label{sec-surface}
Up to now we have assumed that all galaxies of a given apparent
magnitude will be detected.  In practice, a galaxy will be detected only if
its observed surface brightness is above a threshold value
($\mu_{det}$) over a minimum angular area (A$_{min}$).  The values of
$\mu_{det}$ and A$_{min}$ are different for each survey.  The observed
surface brightness profile depends on the intrinsic galaxy profile
convolved with a point spread function.  In the detection process, the
image is often further convolved with a smoothing function.  We use
the formalism developed by Yoshii (1993) to compute the detectability
of galaxies.  The detected magnitude of a galaxy differs from the
total integrated magnitude by $\Delta=m_{det}-m_{tot}=
-2.5{\rm log}(\tilde{G}(\theta_{det})/
\tilde{G}(\theta)))$, where $\tilde{G}(\theta)$ is Yoshii's notation for the 
integrated intensity out to radius $\theta$.  The expression for
$\tilde{G}$ depends on a galaxy's intrinsic profile, its redshift, and
the seeing conditions.  The value of $\theta_{det}$ is fixed for an
aperture magnitude, and for an isophotal magnitude, again depends on
the galaxy's properties.  Thus $\Delta$ is a function of galaxy type,
$M$, $z$, and the detection conditions. To compute Eq.~\ref{nm-eq} and 
\ref{nz-eq} we first
iteratively solve Eq.~\ref{mag-eq} for $M$, then determine whether the galaxy survives
the detection criteria before evaluating the LF.

Table~\ref{SB-rules} lists the surface brightness selection criteria
we adopt for the redshift distributions. We have tried to take from
the literature the parameters used to produce the observed
distributions. However, often the information given in the papers is
incomplete and we have made reasonable guesses for the missing
parameters.  In some cases the redshifts come from several sources
with different selection criteria.  Thus the precise quantitative
effects of the selection criteria should not be taken as absolute, but
we believe that the effects shown are quite representative.

\subsection{Quiescent population}
For our quiescent population we assume that the E and S0 types have an
$\rm r^{1/4}$ profile where the central surface brightness, $\mu_0$ at
$z=0$ is assumed to be 14.8$B_J$ mag arcsec$^{-2}$ (Fish 1964). For the
spirals we assume an exponential profile with $\mu_0$=21.0 for types
Sab and Sc, and $\mu_0$=22.2 for Sdm and Irr (van der Kruit 1987).
These are the mean apparent central surface brightnesses for these
galaxy types, rather than a value corrected for inclination.  This is
appropriate because in our calculation we make no consideration of
galaxies of different inclinations. Van der Kruit's results are fully
consistent with those of Freeman (1970) who obtained $\mu_0$=21.7
averaged over all types after inclination correction.


\subsection{Burst population}
\begin{figure}
{\plotone{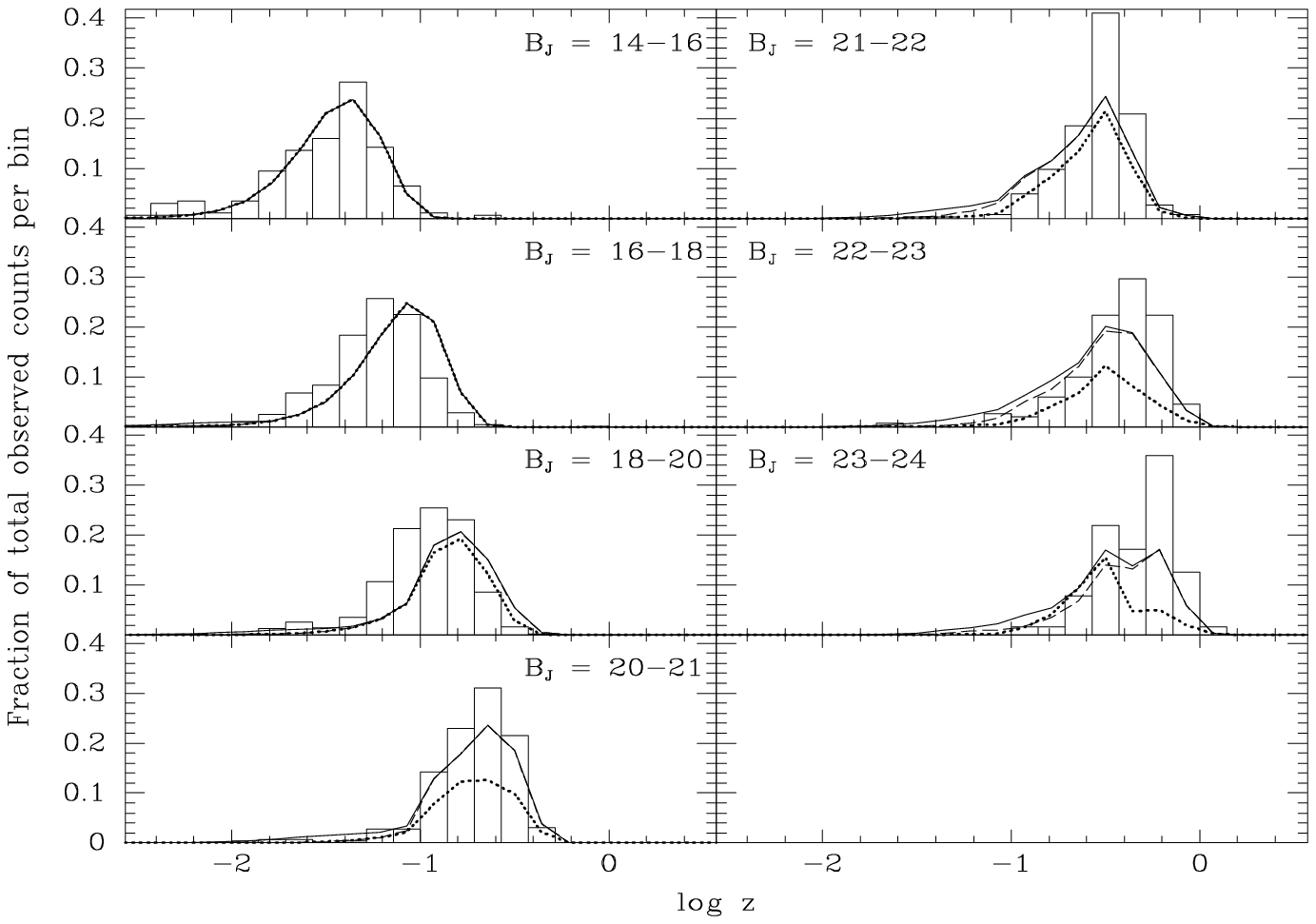}}
\caption[Redshift distributions with SB selection]{\label{sb-nz}}
$B_J$ selected redshift distributions incorporating surface
brightness selection for the baseline plus local IMF
burst populations.  The solid, dashed and dotted curves are for
populations where a $10^8 M_{\sun}$ burst has $r_e$ = 1, 3, and 10 kpc
respectively ($H_0=50\hunits$, $q_0=0.05$) .  The three curves are
degenerate in the upper-left plots because the burst population does
not contribute at those magnitudes.
\end{figure}

For the burst population we make the same assumption with regards
to central surface brightness as for the quiescent 
population: galaxies of the same age have the same central surface
brightness for all magnitudes. Figure~\ref{sb-nz} shows the predicted
redshift distributions for three scalings: for a $10^8 M_{\sun}$ burst
the galaxies have half-light radii ($r_e$) of 1 (solid), 3 (dashed),
and 10\,kpc (dotted). 

In the $B_J=21-23$ bins, the low-redshift tails are strongly dependent
on the sizes of the burst galaxies.  With the normal IMF, a $10^8
M_{\sun}$ burst corresponds to an absolute magnitude of $M_{B_J} =
-18$ during the bright phase.  High-resolution imaging of a sample
of galaxies with $21 < B_J < 22.5$ (Colless et al. 1994) shows that
such a galaxy should have $r_e \approx 2$\,kpc. This corresponds to
between the dashed and solid lines of Figure~\ref{sb-nz}, at a level
where the low-redshift tails are still in conflict with the observations.
It is worth noting that the effects of surface brightness selection are
different in each of the redshift bins. This points out the importance of
understanding the selection criteria for each survey.

\subsection{Low surface brightness galaxies}
\label{lsb-sec}
Disk galaxies with low surface brightness (LSB) are known to exist in
the local universe (McGaugh 1992) but are not included in the
luminosity functions derived from large area photographic surveys
(Impey 1993).  A derivation of the luminosity function of these
galaxies is not yet available but they are believed to exist in
numbers comparable to the high surface brightness galaxies.  McGaugh
(1994) has argued that these LSB galaxies are identical to the faint
blue galaxies that are missing from the baseline model.  We now add
such a population to our model.  The evolutionary history of such
galaxies is not well understood (McGaugh \& Bothun 1994) so for
simplicity we assume a single SED of constant star formation with age
5 Gyr. This SED has colors $U-B=-0.18, B-V=0.33,$ and $V-I=0.96$,
chosen to match the median colors of the LSB galaxies, $U-B=-0.17,
B-V=0.44,$ and $V-I=0.89$ (McGaugh \& Bothun 1994).  We will assume a
single luminosity function identical to that for high surface
brightness galaxies: $\phi=0.00175\rm\,Mpc^{-3}\,mag^{-1}$,
$\alpha=-0.97, M^*_{B_J}=-21.0$ (\cite{Loveday92}).  Additionally we
assume no evolution for the LSB SED. By this we do not imply that
these galaxies do not evolve, but we assume that the population as a
whole does not change with time.  An exponential surface brightness
law with $\mu_0=23.5$ is adopted for all the LSB galaxies.  These
assumptions will certainly be subject to revision as more information
on the LSB galaxy population becomes available.

\begin{figure}
{\plotone{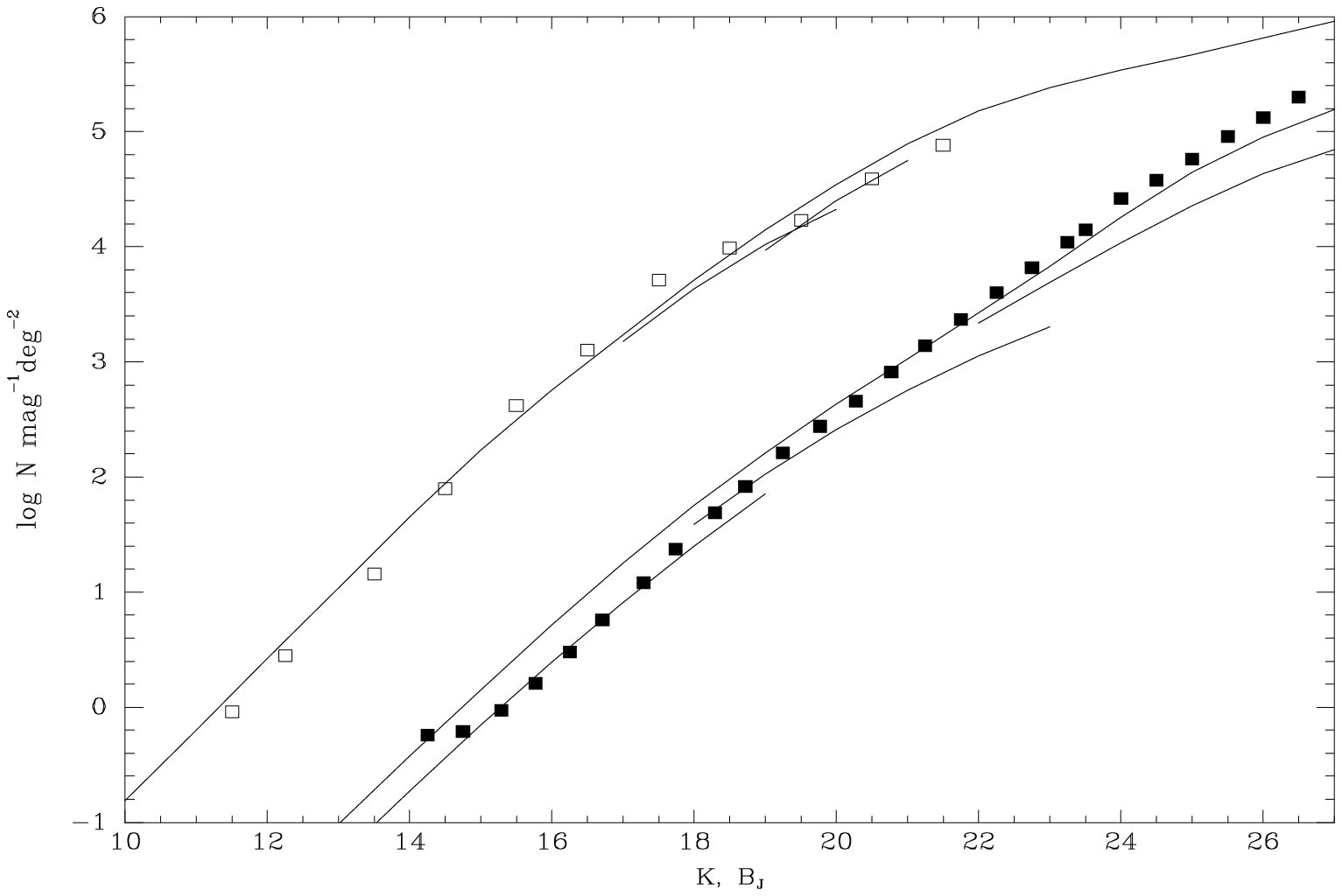}}
\caption[Galaxy counts with LSB galaxies]{\label{LSB-nm}}
Number vs. magnitude for a baseline population plus an equivalent
number of LSB galaxies.  The long-upper curves are for no surface brightness
selection.  The lower 3 curves show what should actually be observed
using the selection criteria that
are listed in Table~\ref{SB-rules}.  The predictions match the observations
fairly well for $B_J\lesssim23$.
\end{figure}

Figure~\ref{LSB-nm} shows that adding the LSB population has its
strongest effect on the $B_J$ counts between 18 and 23. The nearby
galaxies are not affected greatly because the surface brightness
threshold in the local survey (\cite{Maddox90b}) is too high to detect
many of them. For those that are detected, the extrapolation to pseudototal
magnitudes (\cite{Maddox90a}) underestimates the actual magnitude by
up 1 mag.  At intermediate magnitudes the increased sensitivity of
the surveys allows them to be detected.  However, at high redshift in
the absence of evolution, the galaxies once again are too faint to
detect.  The known population of LSB galaxies can explain the
observations only for $B_J \lesssim 23$.  A recent analysis of LSB galaxies by Ferguson \& McGaugh (1995) reaches a similar conclusion.

\section{K-selected colors}
\label{sec-kcolors}
\begin{figure}
{\plotone{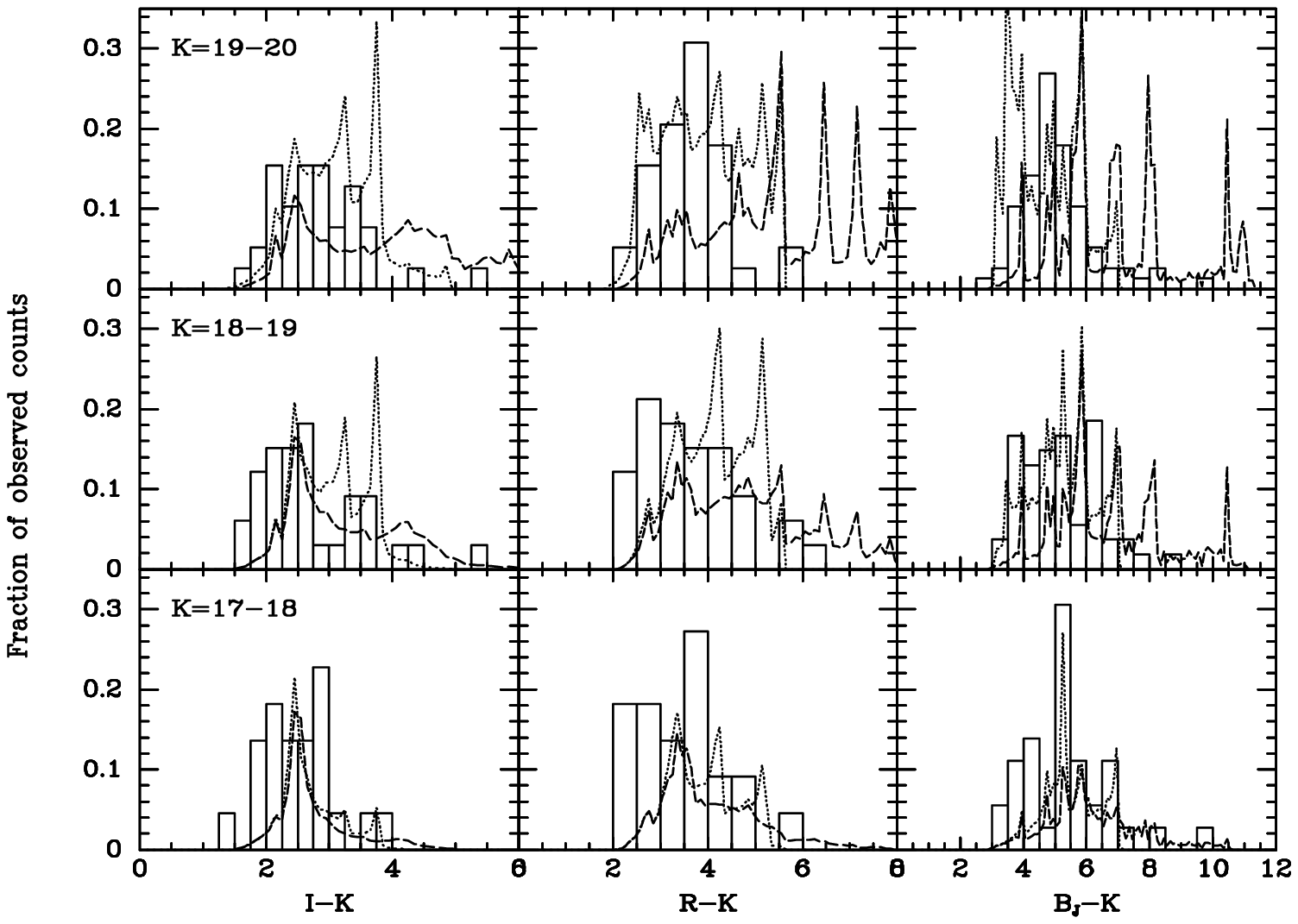}}
\caption[Colors of $K$-selected galaxies]{\label{kcolors}}
Colors of $K$-selected galaxies.  The $K$-magnitude range is shown in the
left panel of each row.  The histogram shows data from 
McLeod et al.\ (1995).
  Dashed line is for the baseline population with constant
SFR.  Dotted line is for the modified population with exponential SFR.
\end{figure}

\begin{figure}
{\plotone{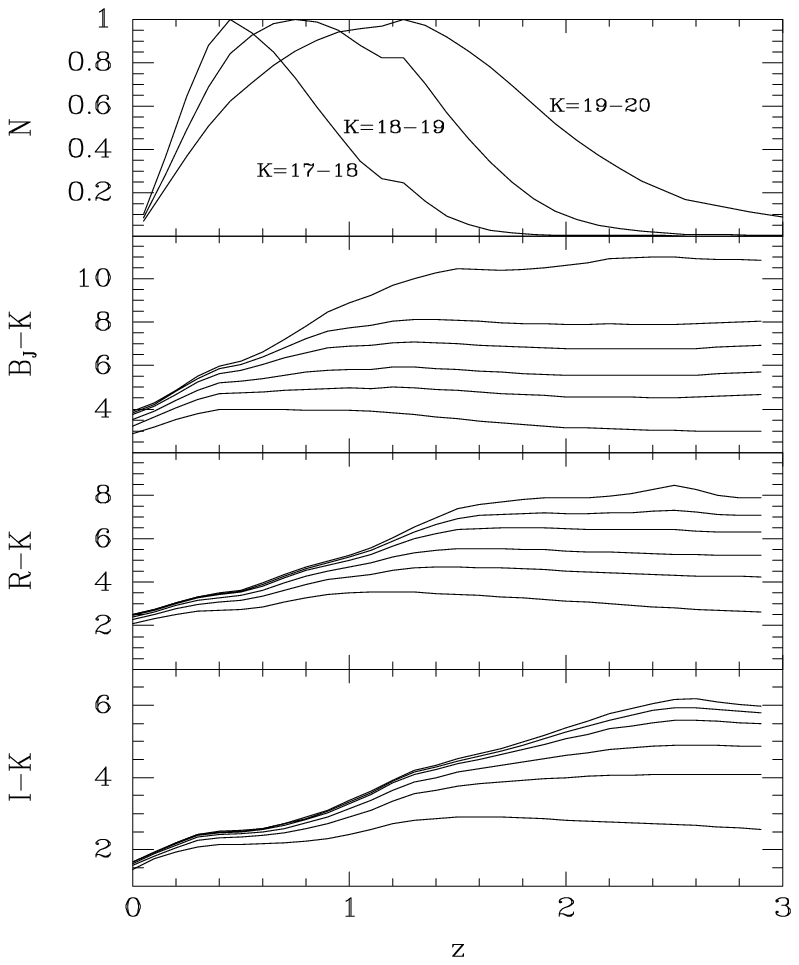}}
\caption[Colors of the baseline model]{\label{kzdist}}
The bottom three panels show the colors of the evolving baseline population as a
function of redshift.  The six curves are for the E though Im types from 
top to bottom.  The upper panel shows the predicted redshift distributions
for a $K-$selected population.  From left to right the three curves are for
$K=17-18$, $18-19$ and $19-20$.
\end{figure}

Galaxies selected at $K-$band are a powerful probe of higher redshifts
because of the smaller KE-corrections at longer wavelengths.
Figure~\ref{kcolors} shows the color distributions of galaxies from
%
the sample described by McLeod et al. (1995). Overplotted (solid lines)
is the predicted distribution from the baseline population.
It is immediately clear that in the $K=19-20$ range, the model predicts
colors that are too red.  From Figure~\ref{kzdist} we
see that the predicted red galaxies are high-redshift objects with low
amounts of star formation.  The spikes in the $R-K$ and $B_J-K$
distributions are caused by the fact the model contains a
discrete number of galaxy types and the color curves are flat at high
redshift.  In reality these distributions will be smoothed out.  A
particularly important color to consider is $I-K$ because even at
redshifts approaching 2, the observed $I$ band is still emitted
longward of 3000\AA~where we have fairly good knowledge of  what zero-redshift
SEDs look like.  The blue colors of the galaxies suggest two
possibilities.  The first is that the galaxies in the $K=19-20$ range
are not at such high redshift.  This would be the case if significant
numbers of galaxy mergers occurred since redshift $1-2$ and we are
seeing the galaxies before merging. 

\begin{figure}
{\plotone{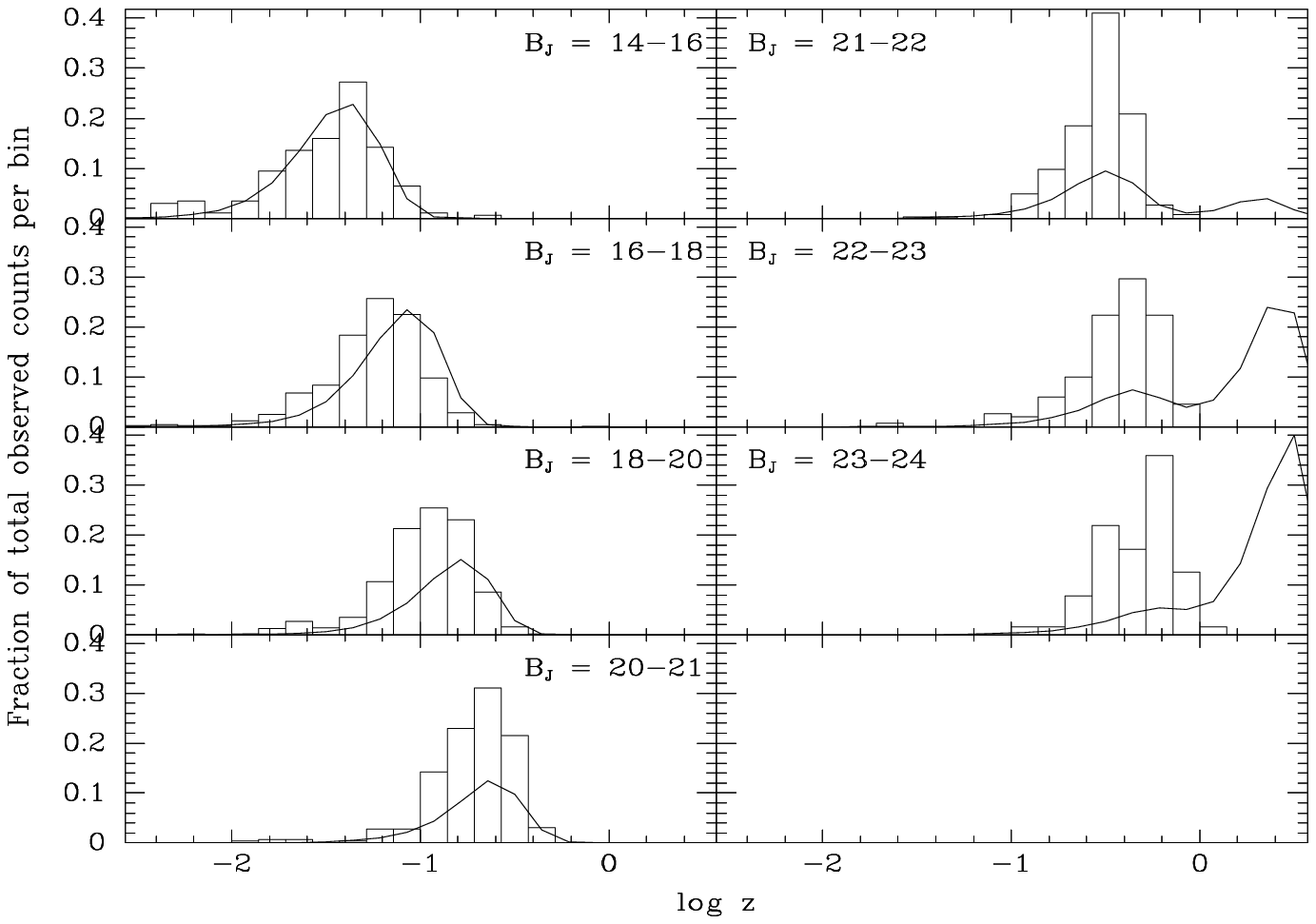}}
\caption[Redshift distributions for the modified baseline populations]
{\label{blue-nz}} $B_J$-selected redshift distributions for the
exponential modified baseline population.  The increased evolution of
the red galaxies adds too many galaxies in the blue.  For this model
to work, dust must extinguish the blue light.
\end{figure}

The second possibility is that the star-formation rate of present-day red
galaxies was significantly higher at $z\approx1$.  We now replace the 
E and S0 populations from the baseline model  with 
an exponentially decaying star-formation rate SED with $\tau=1$Gyr. 
The Sab SED is replaced with $\tau=2$Gyr.  For comparison with
Table~\ref{quiesc-lf} these SEDs have present day
(age = 13.5\,Gyr) $B-V$ colors of 0.91 and 0.87 respectively.
At $z=1.5$ the observed colors are considerably bluer than their
constant-SFR counterparts with $I-K=3.8$ and 3.25 respectively.
The dotted curves in Figure~\ref{kcolors} show a better agreement
with the observations than the baseline model.  However, the 
$B_J$ redshift distributions now predict too many high-redshift
galaxies compared with the observations (Figure~\ref{blue-nz}).

The definitive test to distinguish between merging and an increased
SFR requires determining redshifts for $K$-selected galaxies. Elston
(1994) has reported results of a redshift survey of galaxies with
$K<18$ and $R<22$ that shows a median redshift $>0.5$, implying that
mergers are relatively unimportant at low redshift. In contrast, a
partially complete survey to $K=20$ (Songaila et al. 1994) shows that
the majority of galaxies are not at high redshift.  Additional surveys
to $K=$19--20 are currently underway in the UK and Arizona. The
discrepancy between the baseline model and the observations becomes
most apparent in the $K=19-20$ range, but determining complete samples of
redshifts here will be quite difficult as a majority of the galaxies
have $B_J>24$. The high-redshift galaxies, unfortunately, will be the
most difficult to get redshifts for as the $k$-correction makes them
the reddest.

\section {Summary}
We have explored several scenarios for explaining the magnitude,
redshift and color distributions of galaxies.  The baseline population
of galaxies  consists of an initial burst of
star formation, followed by constant star formation thereafter. We conclude
\begin{enumerate}

\item The baseline population with passive evolution and 
the Shanks (1990) color-dependent luminosity function underestimates
the number of galaxies at faint $B_J$.

\item Adding a population of starburst galaxies to increase the counts
at $B_J$ adds too many low-redshift remnants not seen in the $z$
distributions.  The red stars in the starburst galaxies do not fade
enough to be absent from local surveys.  

\item Changing the IMF of the burst population removes the low-redshift tail
problem. Both starburst populations produce galaxies that are too blue in 
$B_J-R_F$.  Alleviating the blueness by adding dust or old stars would
increase the low-redshift excess problem.

\item Adding surface brightness selection to the simulation
has significant effects on the $n(m)$ and $n(z)$ relations at all
magnitudes as previously discussed by Yoshii (1993).  The low-$z$
tails produced by the local IMF starburst population are diminished
but not eliminated.

\item Adding a population of low surface brightness galaxies, which are known
to exist locally, removes much of the discrepancy in the counts
brighter than $B_J = 23$.  At fainter magnitudes there is still a
problem.  Full understanding of this effect will require LFs corrected
for surface brightness effects.

\item The blue $I-K$ colors of galaxies selected with $K=19-20$ show that 
an old quiescent population is not an adequate model for elliptical
galaxies.  Making the galaxies bluer can be accomplished by having
them at $z$=1 but forming more stars. This appears to be ruled out by
the $B_J$ selected redshift distributions.  The alternative is to
increase the number of lower redshift galaxies. They are bluer because
of the $k$-corrections. The physical mechanism to accomplish this is not
clear, though one possibility is for these galaxies to be in a premerged 
state.

\end{enumerate}

One of the most important next steps in modeling galaxy distributions
will be making use of surface brightness corrected luminosity
functions to model the local population of galaxies.  The analysis of
a LSB population in
\S\ref{lsb-sec} is an important first step, but a more accurate
treatment should be possible soon 
using recently derived luminosity functions for LSB galaxies
(Sprayberry, 1994).
The results of \S\ref{lsb-sec} suggest that at higher redshift a starburst
population will still be necessary.  In the present models we
considered the extreme situation of a given galaxy having only a
single burst in its life.  The result showed that with a normal IMF,
these bursts leave remnants that should be detected nearby but are not
accounted for in the local LF. Using a truncated IMF allows the
galaxies to fade sufficiently. However, these model burst populations
are all too blue in their $B_J\!-\!R_F$ colors implying that the
observed galaxies must also have an older red population.  
Wyse \& Silk (1987) have argued that such bimodal star formation
may have occurred in the solar neighborhood.  Unfortunately, the
presence of an old population exacerbates the lack-of-fading 
problem.  Further analysis will be required to evaluate this scenario.

Closely coupled to the issue of starbursts is that of interactions and
merging (Larson \& Tinsley 1978; Carlberg \& Charlot 1992).
Incorporating merging into the modeling problem is complicated because
an interaction changes the total number of galaxies, the amount of
light (through induced star-formation) and potentially the
scale-lengths of the light.  Yoshii (1993) has suggested that
pre-merger galaxies will not be detected due to surface brightness
selection effects. In particular he argues that premergers cannot
account for the number of galaxies required to make an $\Omega=1$
cosmology fit the $B_J$ and $K$ counts as is suggested by Broadhurst
et al. (1992).

The starburst treatment that we have introduced is a good starting
approximation to increased merging in the past.  Our starburst
galaxies represent the post-merger galaxies.  An useful addition to
the modeling program will be to add the number evolution associated with
merging to the existing code.  This will be especially important in 
investigating the apparent inconsistencies with the passive evolution
model and the colors and redshifts of faint $K$-selected samples.

Another current deficiency is the state of the $n(m)$ data.
Tables~\ref{obsU-tab}--\ref{obsK-tab} are compiled from over two dozen 
separate sources, each with its own detection criteria, which are often
not stated completely.  To make matters more complicated, many different
filters have been used.  With large format CCD mosaics becoming available
it is worth considering redoing large parts of the observations in a 
consistent manner. The ideal survey would image the same areas of the
sky in multiple filters, providing color information for a large number
of galaxies, and reducing discrepancies  introduced by large scale structure.
At the same time, the surface brightness selection criteria could be
done in a consistent way, thus eliminating some of the difficulties in
analyzing published data sets.

\acknowledgments
We would like to thank G. Bruzual for providing his galaxy evolution
software,  C. Gronwall and D. Koo for providing their color and redshift
data, and G. Bernstein for many useful discussions.  This work fulfilled part
of the dissertation requirements of B.A.M. at the University of Arizona and
was supported through an NSF Faculty Award for Women.

\clearpage
\begin{table}
\caption{U-band counts\label{obsU-tab}}
\end{table}

\begin{table}
\caption{B$_J$-band counts\label{obsBj-tab}}
\end{table}

\begin{table}
\caption{r-band counts\label{obsr-tab}}
\end{table}

\begin{table}
\caption{I-band counts\label{obsI-tab}}
\end{table}

\begin{table}
\caption{K-band counts\label{obsK-tab}}
\end{table}

\begin{table}
\caption{Quiescent population luminosity functions\label{quiesc-lf}}
\end{table}

\begin{table}
\caption{Burst characteristics for 0.1--125$M_{\sun}$ IMF\label{burst-lf}}
\end{table}

\begin{table}
\caption{Surface brightness selection criteria\label{SB-rules}}
\end{table}


\clearpage



\end{document}


\begin{planotable}{lrrr}
\singlespace
\tablewidth{15pc}
\tablecaption{U-band counts\label{obsU-tab}}

\tablehead{
\colhead{$U$} &
\colhead{$\log n\,[\rm mag^{-1}deg^{-2}]$} & 
\colhead{$N$} &
\colhead{Ref}
}

\startdata
18.25&	1.76&	31&	1,2\nl
18.75&	1.97&	44&	1,2\nl
19.25&	2.20&	70&	1,2\nl
19.75&	2.53&	150&	1,2\nl
20.25&	2.77&	260&	1,2\nl
20.75&	2.96&	540&	1,2\nl
21.25&	3.22&	420&	1\nl
21.75&	3.57&	920&	1\nl
22.25&	3.68&	170&	3,4\nl
22.75&	4.04&	140&	3,4\nl
23.25&	4.33&	280&	3,4\nl
23.75&	4.54&	450&	3,4\nl
24.25&	4.74&	720&	3,4\nl
24.75&	4.85&	530&	4\nl
\tablerefs{
(1) Koo (1986); (2) Jones et al.\ (1991); (3) Majewski (1989); 
(4) Guhathakurta et al.\ (1990).
}
\end{planotable}







\def\hunits{\hbox{\(\rm \,km\,s^{-1}\,Mpc^{-1}\)}}


\begin{planotable}{lrrr}
\singlespace
\tablewidth{15pc}
\tablecaption{B$_J$-band counts\label{obsBj-tab}}

\tablehead{
\colhead{$B_J$} &
\colhead{$\log n\,[\rm mag^{-1}deg^{-2}]$} & 
\colhead{$N$} &
\colhead{Ref}
}

\startdata
14.25\dotfill&	$-0.24$&	55&	1,2\nl
14.75&	$-0.21$&	60&	1,2\nl
15.29&	$-0.03$&	800&	3\nl
15.77&	0.21&	1400&	3\nl
16.25&	0.48&	2600&	3\nl
16.71&	0.76&	4900&	3\nl
17.29&	1.08&	10000&	3\nl
17.74&	1.37&	20000&	3\nl
18.29&	1.69&	42000&	3\nl
18.72&	1.92&	71000&	3\nl
19.25&	2.21&	140000&	3\nl
19.77&	2.44&	240000&	3\nl
20.27&	2.66&	400000&	3\nl
20.77&	2.91&	690000&	3\nl
21.25&	3.14&	2200&	4\nl
21.75&	3.37&	3800&	4\nl
22.25&	3.60&	6400&	4\nl
22.75&	3.82&	11000&	4\nl
23.25&	4.04&	17500&	4\nl
23.5&	4.15&	2000&	5,6,7\nl
24.0&	4.42&	3800&	5,6,7\nl
24.5&	4.58&	5100&	5,6,7\nl
25.0&	4.76&	2600&	5\nl
25.5&	4.96&	3600&	5,8\nl
26.0&	5.12&	4100&	5,8\nl
26.5&	5.30&	120&	8\nl
\tablerefs{
(1) Ciardullo (1987);
(2) Heydon-Dumbleton et al. (1989);
(3) Maddox et al. (1990);
(4) Jones et al. (1991);
(5) Tyson (1988);
(6) Metcalfe et al. (1991);
(7) Neuschaefer et al. (1991);
(8) Lilly et al. (1991).
}
\end{planotable}






\def\hunits{\hbox{\(\rm \,km\,s^{-1}\,Mpc^{-1}\)}}



\begin{planotable}{lrrr}
\singlespace
\tablewidth{15pc}
\tablecaption{r-band counts\label{obsr-tab}}

\tablehead{
\colhead{$r$} &
\colhead{$\log n\,[\rm mag^{-1}deg^{-2}]$} & 
\colhead{$N$} &
\colhead{Ref}
}

\startdata
15.25&	0.97&	1800&	2,3\nl
15.75&	1.17&	3000&	1,2,3\nl
16.25&	1.40&	5200&	1,2,3\nl
16.75&	1.63&	8800&	1,2,3\nl
17.25&	1.87&	15000&	1,2,3\nl
17.75&	2.12&	27000&	1,2,3\nl
18.25&	2.33&	42000&	2,3\nl
18.75&	2.55&	69000&	2,3\nl
19.25&	2.83&	1400&	1,8\nl
19.75&	3.05&	2400&	1,8\nl
20.25&	3.24&	3700&	1,8\nl
20.75&	3.42&	5600&	1,8\nl
21.25&	3.59&	8300&	1,8\nl
21.75&	3.75&	12000&	1,8\nl
22.25&	3.94&	660&	4,5,6,7\nl
22.75&	4.13&	930&	4,6,7\nl
23.25&	4.30&	450&	4,6\nl
23.75&	4.56&	830&	4,6\nl
24.25&	4.73&	1200&	4,6\nl
24.75&	4.95&	1500&	6\nl
25.25&	5.17&	1900&	6\nl
\tablerefs{
(1) Stevenson et al. (1986);
(2) Sebok (1986);
(3) Picard (1991);
(4) Hall \& Mackay (1984);
(5) Yee et al. (1986);
(6) Tyson (1988);
(7) Metcalfe et al. (1991);
(8) Jones et al. (1991).
}
\end{planotable}





\def\hunits{\hbox{\(\rm \,km\,s^{-1}\,Mpc^{-1}\)}}




\begin{planotable}{lrrr}
\singlespace
\tablewidth{15pc}
\tablecaption{I-band counts\label{obsI-tab}}

\tablehead{
\colhead{$I$} &
\colhead{$\log n\,[\rm mag^{-1}deg^{-2}]$} & 
\colhead{$N$} &
\colhead{Ref}
}

\startdata
17.55&	2.29&	21&	1\nl
18.05&	2.69&	52&	1\nl
18.55&	2.95&	220&	1\nl
19.05&	3.20&	400&	1\nl
19.55&	3.47&	700&	1\nl
20.05&	3.69&	1200&	1\nl
20.55&	3.86&	1800&	1\nl
21.25&	3.93&	320&	2,3\nl
21.75&	4.13&	500&	2,3\nl
22.25&	4.29&	440&	4,2\nl
22.75&	4.43&	610&	4,2\nl
23.25&	4.59&	850&	4,2\nl
23.75&	4.81&	1200&	2,5\nl
24.25&	4.98&	1700&	2,5\nl
24.75&	5.04&	31&	5\nl
\tablerefs{
(1) Koo (1986);
(2) Tyson (1988);
(3) Hintzen et al (1991);
(4) Hall \& Mackay (1984);
(5) Lilly et al. (1991).
}
\end{planotable}




\def\hunits{\hbox{\(\rm \,km\,s^{-1}\,Mpc^{-1}\)}}





\begin{planotable}{lrrr}
\singlespace
\tablewidth{15pc}
\tablecaption{K-band counts\label{obsK-tab}}

\tablehead{
\colhead{$K$} &
\colhead{$\log n\,[\rm mag^{-1}deg^{-2}]$} & 
\colhead{$N$} &
\colhead{Ref}
}

\startdata
11.5&	$-0.04$&	32&	1\nl
12.25&	0.45&	49&	1\nl
13.5&	1.16&	23&	2\nl
14.5&	1.90&	146&	2,3\nl
15.5&	2.62&	155&	2,3\nl
16.5&	3.10&	491&	2,3\nl
17.5&	3.71&	232&	2,4\nl
18.5&	3.99&	207&	2,4\nl
19.5&	4.23&	158&	2,4\nl
20.5&	4.59&	104&	2,4\nl
21.5&	4.88&	74&	2,4\nl
\tablerefs{
(1) Mobasher et al. (1986);
(2) Gardner et al (1993);
(3) Glazebrook et al (1993);
(4) McLeod et al (1994).
}
\end{planotable}



\def\hunits{\hbox{\(\rm \,km\,s^{-1}\,Mpc^{-1}\)}}






\begin{planotable}{lllcrr}
\singlespace
\tablewidth{0pc}
\tablecaption{Quiescent population luminosity functions\label{quiesc-lf}
}
\tablehead{
Type&SFR\tablenotemark{a} & \bv & \multicolumn{3}{c}{Schechter parameters
\tablenotemark{b}} \nl
    &                     &     & $\phi^*[{\rm Mpc^{-3}\,mag^{-1}]}$
 & $M^*_{B_J}$ & $\alpha$}
\startdata
E\dotfill&0.00001&0.94&$1.92\times10^{-4}$&$-21.1$&$-0.7$\nl
S0\dotfill&0.001&0.90&$1.92\times10^{-4}$&$-21.1$&$-0.7$\nl
Sab\dotfill&0.003&0.83&$1.92\times10^{-4}$&$-21.1$&$-0.7$\nl
Sc\dotfill&0.01&0.7&$2.60\times10^{-4}$&$-21.4$&$-1.1$\nl
Sdm\dotfill&0.03&0.57&$0.85\times10^{-4}$&$-21.5$&$-1.5$\nl
Im\dotfill&1.0&0.42&$0.85\times10^{-4}$&$-21.5$&$-1.5$\nl
\tablenotetext{a}{Constant SFR Gyr$^{-1}$ relative to initial burst}
\tablenotetext{b}{Shanks (1990); $H_0=50\hunits$}
\tablenotetext{}{} 
\tablenotetext{}{}
\tablenotetext{}{}
\tablenotetext{}{}
\tablenotetext{}{}
\tablenotetext{}{}
\tablenotetext{}{}
\tablenotetext{}{}
\tablenotetext{}{}
\end{planotable}


\def\hunits{\hbox{\(\rm \,km\,s^{-1}\,Mpc^{-1}\)}}







\begin{planotable}{lcll}
\singlespace
\tablewidth{0pc}
\tablecaption{Burst characteristics for 0.1--125$M_{\sun}$ IMF\label{burst-lf}}
\tablehead{
\makebox[3cm]{}  &Present   &                             &                      \nl
Redshift&Age (Gyr) & $\Delta M\tablenotemark{a}$ &$\mu$\tablenotemark{b}
}
\startdata
0.10-0.14\dotfill& 2.0 & 4.3 & 26.1 \nl
0.14-0.19\dotfill& 2.6 & 4.7 & 26.4 \nl
0.19-0.27\dotfill& 3.4 & 5.0 & 26.7 \nl
0.27-0.37\dotfill& 4.3 & 5.2 & 26.9 \nl
0.37-0.51\dotfill& 5.5 & 5.5 & 27.2 \nl
0.51-0.72\dotfill& 6.7 & 5.6 & 27.3 \nl
0.72-1.00\dotfill& 8.0 & 5.7 & 27.4 \nl
\tablenotetext{a}{Amount of fading between peak brightness and present.}
\tablenotetext{b}{Present day central surface brightness so that 
$10^8 M_{\sun}$ burst has $r_e = 3$kpc.}
\end{planotable}

\def\hunits{\hbox{\(\rm \,km\,s^{-1}\,Mpc^{-1}\)}}








\begin{planotable}{lccccll}
\tablecaption{Surface brightness selection criteria\label{SB-rules}}
\tablewidth{0pc}
\singlespace
\tablehead{
Filter &  m &  FWHM & $\mu_{det}$ &$r_{min}$ & Magnitude & Source\nl
       &    &  $[\arcsec]$&$[\rm mag/\Box\arcsec]$&$[\arcsec]$& Type &
}
\startdata
\multicolumn{7}{c}{Criteria for $n(z)$}\nl
\hline
\nl
$B_J$ &  14-18 & 1.0\tablenotemark{a} & 25.0 & 2.0\tablenotemark{a} & Total & Metcalfe et al. (1989)\nl
$B_J$ &  18-21 & 1.8\tablenotemark{a} & 25.5\tablenotemark{a} & 0.6\tablenotemark{a} & Total\tablenotemark{a}&Koo \& Kron (1992) \nl
$B_J$ &  21-23 & 1.8 & 26.5 & 0.6 & Total & Colless et al. (1990)  \nl
$B_J$ &  23-24 & 1.1\tablenotemark{a} & 29.0 & 0.8\tablenotemark{a} & 3\arcsec\ aperture&Lilly et al. (1991) \nl\nl
\hline
\multicolumn{7}{c}{Criteria for $n(m)$}\nl
\hline
\nl

$B_J$& $<19$ & 1.0 & 24.7 & 1.1 & Pseudototal\tablenotemark{b} & Maddox et al. (1990b)\nl
$B_J$& 18-23 & 1.8 & 26.5 & 0.6 & Isophotal &Jones et al. (1991)\nl
$B_J$& 22-27 & 1.0 & 29.0 & 0.8 & Isophotal &Tyson (1988)\nl
$K$  & 17-20 & 2.1 & 23.0 & 1.3 & Pseudototal\tablenotemark{c} & McLeod et al. (1994)\nl

$K$ & 19-21 & 1.8 & 23.6 & 0.5 & 3\arcsec\ aperture& Cowie\tablenotemark{d}\nl
\tablenotetext{a}{Estimated--not given in source.}
\tablenotetext{b}{Used correction scheme given in Maddox et al. (1990a)}
\tablenotetext{c}{Flux is measured in area double that defined by $\mu_{det}$}
\tablenotetext{d}{As presented by Yoshii (1993)}
\end{planotable}